\documentclass[a4paper,twoside]{article}

\usepackage{epsfig}
\usepackage{subcaption}
\usepackage{calc}
\usepackage{amssymb}
\usepackage{amstext}
\usepackage{amsmath}
\usepackage{amsthm}
\usepackage{multicol}
\usepackage{pslatex}
\usepackage{apalike}
\usepackage{algorithm2e}
\usepackage[bottom]{footmisc}
\usepackage{SCITEPRESS}     

\begin{document}

\title{Personalized Recommendation Systems using Multimodal, Autonomous, Multi Agent Systems}

\author{\authorname{Param Thakkar\sup{1}\orcidAuthor{0009-0009-1298-6851}, Anushka Yadav\sup{1}\orcidAuthor{0009- 0009-0454-0171}}
\affiliation{\sup{1}Veermata Jijabai Technological Institute, Matunga, Mumbai, India - 400019}
\email{puthakkar\_b22@ce.vjti.ac.in, asyadav\_b22@ce.vjti.ac.in}
}

\keywords{Product Recommendation Systems, Multi Modal Systems, Large Language Models, Autonomous AI Agents, Personalized Product Recommendations}
\abstract{This paper describes a highly developed personalised recommendation system using multimodal, autonomous, multi-agent systems. The system focuses on the incorporation of futuristic AI tech and LLMs like Gemini-1.5-pro and LLaMA-70B to improve customer service experiences especially within e-commerce. Our approach uses multi agent, multimodal systems to provide best possible recommendations to its users. The system is made up of three agents as a whole. The first agent recommends products appropriate for answering the given question, while the second asks follow-up questions based on images that belong to these recommended products and is followed up with an autonomous search by the third agent. It also features a real-time data fetch, user preferences-based recommendations and is adaptive learning. During complicated queries the application processes with Symphony, and uses the Groq API to answer quickly with low response times. It uses a multimodal way to utilize text and images comprehensively, so as to optimize product recommendation and customer interaction.}

\onecolumn \maketitle \normalsize \setcounter{footnote}{0} \vfill

\section{\uppercase{Introduction}}
\label{sec:introduction}
 For the past few years, the burgeoning growth of AI and Machine Learning technologies has defined how e-commerce and the customers tend to these frameworks. This establishment introduces a unique system that employs advantageous AI frameworks and large language models in improving user experiences of the mentioned models\cite{wang2024userbehaviorsimulationlarge}. The system uses multi agent systems for recommendations leveraging the capabilities of large language models such as Gemini-1.5-pro and LLaMA-70B. These sophisticated language services work collaboratively and on demand execute internet queries and obtain information in order to furnish users with various available products\cite{wu2024mas4poimultiagentscollaborationpoi}.
 
The foundation of the system contains three AI agents that focus on core functions including product recommendation, answering questions based on images, and market trend analysis: the product suggestion agent suggests a product to the consumer based on his preferences and trends as well as the internal trends of the site and third party sites. The product recommender agent gives the most suitable brands and goods in regard to user specifications and the current trends. The image model, on the other hand, processes pictures and gives suggestions based on contextual information from the pictures. The market analysis agent, provides recommendations by analyzing various trends and patterns in the current markets and how relevant are the items that are being recommended.

In order to exhibit the maximum possible performance, the application employs Groq API in performing the Large Processing Unit (LPU) inference that cuts down the response which would last several minutes to a few milliseconds. This allows a smooth conversation for users, even while undertaking such challenging AI-related tasks, because there is no lag.

Next, the system will like all prospects and returning patrons of the system to create profiles indicating their preferences, purchases made and items they have browsed to improve on recommendations made in the system. Reinforcement learning and other adaptive learning technologies will help the system learn and develop over time, hence changing how things work. These improvements on the technology as noted earlier will revolutionize the e-commerce sector in helping consumers enjoy personalized shopping, price comparison and visual search features\cite{Guo2010E-commerce}, \cite{Almahmood2022Issues}. Moreover, regarding customer support, the application can fully automate query resolution via chatbot and offer personalized recommendations based on public interest in a particular product.

\section{\uppercase{Related Work}}
\subsection{AI agents overview}
AI agents are autonomous entities designed to perceive their environment, reason, and perform actions to achieve specific goals. They can operate independently or collaboratively with other agents or systems. In the context of modern technology, AI agents are integral to numerous applications, from personal assistants like Siri and Alexa to complex autonomous systems in robotics and finance. By mimicking human decision-making, these agents can perform tasks efficiently, scaling processes across multiple domains\cite{10140756}.

\subsection{Importance of AI Agents in Technology}
The importance of AI agents in technology cannot be overstated. They provide automation, intelligence, and adaptability to complex environments, helping to solve tasks that require quick decisions or real-time responses. In sectors such as e-commerce, healthcare, and cybersecurity\cite{10.1145/336992.337035}. AI agents enhance efficiency and provide personalized services by interpreting vast amounts of data and delivering actionable insights\cite{inproceedings}. Their adaptability to evolving data makes them indispensable in dynamic environments where human intervention would be slow or less precise\cite{li2024incorporatingexternalknowledgegoal}.

\subsection{LLM-Based Agents for Recommendations}
Large Language Models (LLMs) such as OpenAI's ChatGPT, Gemini, and Meta's Llama have revolutionized the implementation of AI agents in recommendation systems\cite{Wang2023RecMind:}. These models, trained on vast amounts of data, possess strong capabilities in natural language understanding, contextual learning, and decision-making, enabling them to operate autonomously across varied contexts \cite{brown2020languagemodelsfewshotlearners}. In recommendation scenarios, LLM-based agents can analyze user inputs, predict preferences, and suggest personalized recommendations dynamically without the need for extensive pre-training on specific datasets\cite{lin2024recommendersystemsbenefitlarge}.

\subsection{Research in LLM-Based Agent Recommendations}
Current research in LLM-based agents for recommendations primarily follows two orientations:

\begin{itemize}
    \item Simulation-Oriented Approaches: These approaches aim to simulate user behaviors or interactions within a specific environment. The primary goal is to mimic real-world user scenarios, enabling the agents to predict actions and recommend products based on simulated user personas. These systems are often used to test hypothetical interactions and optimize recommendation algorithms in controlled environments.
    \cite{wang2024multiagentcollaborationframeworkrecommender}
    \item Recommender-Oriented Approaches: These approaches focus directly on refining the algorithms and architectures behind recommender systems. LLM-based agents in this domain analyze user data, historical purchases, browsing patterns, and preferences to offer precise recommendations in real-time. Unlike simulation-oriented methods, these agents are typically trained on actual user data, and their primary objective is to provide optimized suggestions\cite{wang2024multiagentcollaborationframeworkrecommender}.
\end{itemize}

Our recommendation system stands apart by being both multimodal and scenario-agnostic. Unlike many existing solutions that rely solely on text data or past browsing history, our system integrates multiple input types, including text, images, and real-time market analysis results, making it more versatile and applicable across various use cases. It combines these modalities to deliver a more comprehensive and adaptive recommendation system. Moreover, it doesn't confine itself to specific scenarios or datasets, allowing for a broader range of applications—from product recommendation to image-based query resolution and beyond.

\subsection{Multi-Agent Systems for Recommendations}
A Multi-Agent System (MAS) consists of multiple autonomous AI agents that interact within an environment to achieve collective or individual goals. Each agent in a MAS has its own objectives and decision-making abilities, but they work together through cooperation, negotiation, or competition. In MAS, agents can either be homogeneous (performing the same type of tasks) or heterogeneous (with different specializations and roles).

In recommendation systems, MAS can be employed to improve decision-making by distributing tasks among specialized agents. Each agent can focus on different aspects of the recommendation process. For example:

One agent might focus on analyzing historical user data.
Another agent might specialize in real-time web searches for product availability and pricing.
A third agent could handle user feedback or image recognition to refine suggestions further.

By working together, these agents improve the system's overall efficiency and provide more accurate, personalized recommendations. The decentralized nature of MAS allows for parallel processing, which can handle the complexity of real-time decision-making across multiple data streams.

\subsection{Frameworks for Building Multi-Agent Systems}
Several frameworks and libraries have been developed to support the building and deployment of multi-agent systems:

\begin{itemize}
    \item CrewAI: A modern framework focused on multi-agent systems, CrewAI supports AI-driven agents that interact with each other and external systems, particularly useful in environments like e-commerce and customer support.
    \item Langchain: This framework simplifies the development of LLM-based agents, supporting autonomous decision-making processes, allowing agents to fetch data, analyze it, and make recommendations in real-time.
\end{itemize}

\subsection{Research in Multi-Agent Systems for Recommendations}
Research on MAS in recommendation systems has grown in recent years, focusing on how agents can cooperate to provide more personalized and accurate suggestions\cite{Lee2002Intelligent}. Key research trends include:

\begin{itemize}
    \item Collaborative Filtering with MAS: Multi-agent systems are used to enhance collaborative filtering by having different agents evaluate user similarities, preferences, and trends in parallel, allowing for faster and more precise recommendations.
    \item Hybrid Recommendation Systems: Some MAS research involves integrating content-based and collaborative filtering methods into a hybrid system where different agents specialize in different recommendation techniques.
    \item Task-Oriented Agent Cooperation: Researchers are exploring how agents can specialize in sub-tasks such as text analysis, sentiment detection, and product comparison, making the overall system more robust and scalable.
\end{itemize}

Our system adopts a MAS approach to recommendation, where different agents (product recommendation, image-based QA, and market-trend analysis) specialize in various tasks. This multi-agent framework allows for the efficient processing of diverse inputs (e.g., text, images) and enables the system to dynamically adapt recommendations in real-time across a wide range of scenarios.

By leveraging both LLM-based agents and multi-agent systems, our solution offers a more comprehensive, multimodal approach to product recommendation and customer support, distinguishing it from traditional text-based or single-agent systems\cite{Huang2019Multimodal}.

\begin{figure*}[h!]
    \centering
    \includegraphics[width=0.7\textwidth]{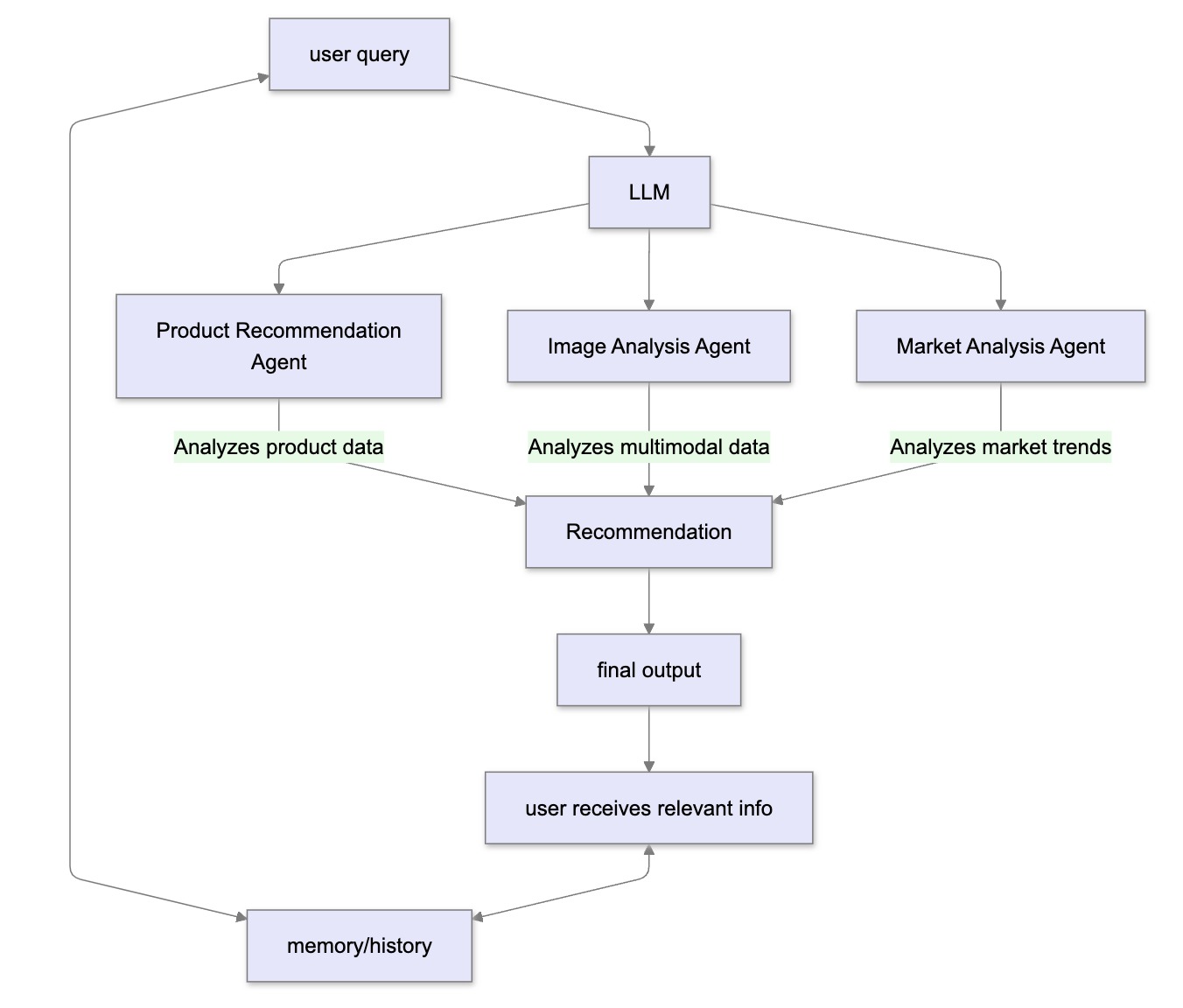}
    \caption{System Design of the Multiagent System}
    \label{fig:example}
\end{figure*}
\section{Methodology}
Our system is capable of providing the best possible and personalised recommendations to users. It does this by using various methods. We employ a user centric, multi agent based system design with dynamic information letting which means that the system will always provide personalised recommendations using multiple agents that are solving various recommendation tasks autonomously and any information required is fetched dynamically during agent workflow runtime\cite{Vullam2023Multi-Agent}.
\subsection{Methods}
\begin{enumerate}
    \item Multi-Agent System Design: This type of application comprises more than one agent which acts at the same time thus a task can be executed in parallel.

    \item Dynamic Information Letting: The system is able to search and obtain information and data on the web on its own and thus provide real time information and recommend products without the use of any pre-trained sets.

    \item User-Centric Design: Focuses on enhancing users’ experiences through personalization and usage of user models.
\end{enumerate}

\subsubsection{Models Used}
The models we used for this system were llama-70b-8192, and gemini-1.5-pro. Though other models can also be used and we have calculated to performance of each one of them as recommenders in the results section below but we majorly focused on working with these two models

\begin{enumerate}
    \item Gemini-1.5-pro: A professional model for the purposes of intricate and sophisticated natural language processing and generation tasks.
    \item LLaMA-70B: A high capacity model which effectively interprets context and retrieves relevant information.
\end{enumerate}
\subsubsection{Agent Roles}

\begin{enumerate}
    \item Product Recommendation Agent: The product recommendation agent focuses on providing the best possible product recommendations to user queries. This agent has been given various tasks to perform in order to give the best recommendations. The tasks it performs are giving the best possible product recommendations on a given user query input. it uses tools used for searching and scraping the entire web to get the relevant information. It analyses this information and gives personalised recommendations according to user prompts and queries.

    \item Multi-modal Recommendation Agent: The multi-modal recommendation agent recommends products based on multi modal data like images. Many times it happens that a person sees a product somewhere and suddenly likes it too much but is unaware of the brand it belongs to. The user in this case can give a picture or a video of the product to the multimodal agent and it finds out similar products from all over the internet for the user and answers user queries based on the particular product along with providing insights and suggestions

    \item Market analysis agent: Gives the most recent market trends. It gives insights and analysis of the recent market trends to provide the best possible tailored recommendations. After keeping in mind the user preferences, it also searches for the recent trends in the market, so that the final recommended product is aligned with both the recent trends as well as the user

\subsubsection{Making Recommendations Based on Multi-Modal Data}
The agents make use of multi-modal data in fulfilling their tasks:

\begin{enumerate}
    \item Text and Images Combination: The image question answering model answers questions by referring to visual data in addition to text information.
    \item Dispersion of data from individual agents: The recommendations made by one agent can benefit from the outcomes of another helpful agent in a unified recommendation approach.
\end{enumerate}

This structured approach allows for an efficient and user-friendly experience, adapting to individual needs and providing timely, relevant product recommendations.

\subsubsection{Technological Design of A Multi-Agent System}
This system makes use of a multi-agent architecture whereby different agents are able to independently and also act collaboration to achieve set goals:

\begin{enumerate}
    \item Agent Architecture: Individual agents are assigned a particular function, making the process efficient.
    \item Communication: Agents provide additional meanings and information resulting in the enhancement of the responding quality above the optimal level.
\end{enumerate}

\subsection{Nature of the Multi-Agent System}
\begin{enumerate}
    \item Task Distribution: System recognizes the queries raised by the user and forwards them to the proper agent corresponding to the type of the task.
    \item Parallel Processing: Several agents are active effectively inhabiting the center of attention to a ratio of time spent to the provision of responses.
    \item Collective Intelligence: The individual agents pass on their relevant output to one CPU which uses the work for the benefit of the user.
    \item Multi-Modal Data Usage : 
    The application utilizes multi-modal data in the interactions thereby also focusing on giving user recommendations:
    \item Textual Data: This includes queries made by users, product description and rating feedback.
    \item Visual Data: The image question answering model takes in images straining the users from the normal usage and answers questions pinning products.
    \item Market Analysis Data: The agents accumulate little or much information concerning the trends in the recent market and analyzes it.
\end{enumerate}

\subsubsection{Forms of Data Used}
We created our own dataset of prompts and relevant outputs to train as well as test our models. This data was obtained and verified for correctness by a human annotator and data collector.

\begin{enumerate}
    \item User Interaction Data: This data is obtained from users through their inputs and their preferences.
    \item Product Data: E-commerce data which includes product description products, reviews and images of the products.
    \item Web Data: Data is also received from the sites in real time through web search.
\end{enumerate}

\subsubsection{Features of used Multi Agent System }
\begin{enumerate}
    \item Generating Personalized Recommendations : The multi-agent system creates personalized recommendations by the multi pronged approach.
    \item User Profiles: User profiles enable the system to comprehend the uniques of an individual and the recommendations they prefer.
    \item Agent Collaboration: The agents utilize the knowledge from user interactions and therefore make recommendations that are accurate.\cite{zhou2020topicguidedconversationalrecommender}
    \item Dynamic Data Analysis: User recommendation is based on the analysis of actively changing trends and users rather than the history.
\end{enumerate}

\subsubsection{Collaborative Task Planning}
A crew of agents is able to plan the execution of more complex tasks made up of a number of subtasks and to carry out each one of them. Each agent puts in its own tools, working on its portion and ensuring that everything is done seamlessly. This multi-agent planning allows the system to perform complex activities with the least amount of human assistance.

\subsubsection{Tools Used for Fetching Relevant Results}
\begin{enumerate}
    \item Groq API: It helps in faster LPU inference which lowers the response time of the system by utilizing this API.
    \item Web Searching Tools: Although data retrieval involves offline evaluation as well, for the assessment via web channels this is exclusively done.
    \item Langchain: It serves as a conductor of different agents and models in the processes of Interactions.
\end{enumerate}

\end{enumerate}
\begin{table*}[hbt!] 
\centering
\caption{Metrics for Product Recommendation Agent}
\begin{tabular}{|c|c|c|c|c|c|}
\hline
\textbf{Models} & \textbf{Precision@K} & \textbf{Recall@K} & \textbf{F-score} & \textbf{MRR} & \textbf{NDCG} \\
\hline
llama3-8b-7192 & 0.6 & 0.3 & 0.4 & 0.33 & 0.34 \\
\hline
llama3-70b-8192 & 0.5 & 1.0 & 0.6667 & 0.4566 & 1.0 \\
\hline
gemma-7b-it & 0.8 & 0.4 & 0.533 & 0.494 & 1.0 \\
\hline
gemma2-9b-it & 0.6 & 0.3 & 0.4 & 1.0 & 0.7103 \\
\hline
\end{tabular}
\end{table*}

\begin{table*}[hbt!]
\centering
\caption{Metrics for Image Analysis Agent}
\begin{tabular}{|c|c|c|c|c|c|}
\hline
\textbf{Models} & \textbf{Precision@K} & \textbf{Recall@K} & \textbf{F-score} & \textbf{MRR} & \textbf{NDCG} \\
\hline
llama3-8b-7192  & 1.0   & 0.5    & 0.667  & 0.228     & 1.0 \\
\hline
llama3-70b-8192 & 0.1   & 0.1    & 0.1333 & 0.5       & 1.0 \\
\hline
gemma-7b-it     & 0.6   & 0.3    & 0.4    & 0.4722    & 1.0 \\
\hline
gemma2-9b-it    & 0.6   & 0.6    & 0.6    & 0.583     & 0.833 \\
\hline
\end{tabular}
\end{table*}
\section{Metrics}
While evaluating our multi agent system and agentic workflows on a variety of real world scenarios we used 6 metrics for quantifying the authenticity, accuracy and trustworthiness of the system \cite{wang2024multiagentcollaborationframeworkrecommender}. These metrics are as follows

\begin{enumerate}
    \item \textbf{Precision at K} \\ \\ This is a common metric for evaluating recommendation systems. It measures the ratio of relevant items to the top K recommended items. It is expressed by the following formula : 
    \\
    \[
        P@K = \frac{\text{Number of Relevant Items in Top K}}{K}
    \]
    \\
    here we consider the number of relevant items in the top K items. K is the total number of items under consideration.
    \\
    \item \textbf{Recall at K}
    \\ \\
    Recall at K evaluates the fraction of relevant items that appear within the top K recommendations, relative to the total number of relevant items in the dataset. In other words, it reflects how many relevant items were correctly retrieved from the dataset in the top K results.
    \\
    \[
        \text{Recall}@K = \frac{\text{Number of Relevant Items in Top K}}{\text{Total Number of Relevant Items}}
    \]
    \item \textbf{F - score}
    \\ \\
    F - score is the harmonic mean of precision and recall, emphasising on the balance between precision and recall scores. It is given by : 
    \[
        F_{\beta} = \frac{(1 + \beta^2) \cdot \text{Precision} \cdot \text{Recall}}{\beta^2 \cdot \text{Precision} + \text{Recall}}
    \]

    where \( \beta \) is a parameter that controls the balance between precision and recall:
    \begin{itemize}
        \item If \( \beta = 1 \), precision and recall are weighted equally.
        \item If \( \beta > 1 \), more weight is given to recall.
        \item If \( \beta < 1 \), more weight is given to precision.
    \end{itemize}
    a high F - score suggests that the system is accurate and comprehensive while a low F - score suggests that the system needs improvement.
    \\
    \item \textbf{Mean Reciprocal Rank (MRR)}
    \\ \\
    Mean Reciprocal Rank (MRR) measures the ranking quality of the first relevant item retrieved for a set of queries.

    It is the reciprocal of the rank for the first relevant item for a particular query. The Mean Reciprocal Rank (MRR) is the average of the reciprocal ranks for a given set of queries. It is expressed as : 
    \\ \\ 
    
    \[
        \text{MRR} = \frac{1}{|Q|} \sum_{i=1}^{|Q|} \frac{1}{\text{rank}_i}
    \]
    \\
    \item \textbf{Normalized Discounted Cumulative Gain (NDCG)}
    \\ \\
    Normalized Discounted Cumulative Gain is a metric used for the evaluation of quality of recommendation systems.

    NDCG is the ratio of Discounted Cumlative Gain (DCG) and Ideal Discounted Cumulative Gain (IDCG)

    Discounted Cumulative Gain (DCG) is a measure of usefulness of a ranked list of items based on their relevance scores. It takes into account the position of the items in the list, giving more emphasis on items appearing earlier. It is given by : 
    \[
        DCG@K = \sum_{i=1}^{K} \frac{rel_i}{\log_2(i + 1)}
    \]
    where:
    \begin{itemize}
        \item \( rel_i \) is the relevance score of the item at rank \( i \).
        \item \( i \) is the rank position, starting from 1.
    \end{itemize}
    Ideal Discounted Cumulative Gain is the highest possible value of Discounted Cumulative Gain (DCG) which is achievable. It is calculated in a way similar to DCG but ordering them in the best possible order.
    It is given by : 
    \[
        IDCG@K = \sum_{i=1}^{K} \frac{rel^{*}_i}{\log_2(i + 1)}
    \]

    where:
    \begin{itemize}
        \item \( rel^{*}_i \) is the relevance score of the items in the ideal ranking.
    \end{itemize}

    Normalized Discounted Cumulative Gain (NDCG) is the ratio of DCG and IDCG. It allows for a value between 0 and 1, making it easier to evaluate and compare different ranking systems. It is given by: 
    \[
        \text{NDCG}@K = \frac{DCG@K}{IDCG@K}
    \]
    NDCG ranges from 0 to 1, with 1 representing the perfect ranking where the most relevant items are ranked at the top of the list.

    \item \textbf{Recall-Oriented Understudy for Gisting Evaluation (ROUGE) Score} 
    \\ \\
    The ROUGE (Recall-Oriented Understudy for Gisting Evaluation) metric is widely used in the domain of automatic text summarization. Despite being developed for the evaluation of machine-generated summaries, the ROUGE score, or Recall-Oriented Understudy for Gisting Evaluation score, has found applications in other fields. ROUGE is a scoring procedure for comparing a summary made by a human or a machine with an ideal summary. The rediscovery procedure, as indicated by the name, involves retrieving stored information. RAP focuses on stating that information directly after asking for it. The experimental type corrected cosine similarity and redundancy elimination. Recall-Oriented Understudy for Gisting Evaluation, also known as ROUGE, was originally designed for evaluating outcomes or machine-generated text in summary form, but it has found several applications in other areas as well. The use of ROUGE scores for testing the retrieval model’s applicability must be considered appropriately. Summarization by generative models of text is non-traditional and scarcely used in authentic texts.
    \\ \\
    The general formula for the ROUGE score is:
    \[
        \text{ROUGE} = \sum_{n=1}^{N} \text{Recall of } n\text{-grams}
    \]
    
    In this formula, the recall of n-grams is defined as the ratio of n-grams present in both the machine-generated and reference summaries to the total number of n-grams in the reference summaries. ROUGE measures the extent to which the machine-generated summary captures important content by reflecting how many of the reference n-grams are included.
    
    For this research, we focused on ROUGE-1, which captures the overlap of single words; ROUGE-2, which captures the overlap of two-word sequences; and ROUGE-L, which measures the longest common subsequence (LCS) between the generated and reference texts. Each of these variations offers different insights into the quality of the summarization.

\end{enumerate}

\section{Results}

We compiled our entire results by considering a variety of scenarios when it comes to recommendation systems. We also tested its answers and their authenticity and accuracy by giving it queries related to a variety of real world use cases of recommendation systems

To compile any of these results, we haven’t used any local gpus or local models. The models we tested were using the groq API key. Due to the rate limits of groq API key, we weren’t able to test the entire agentic workflow of these systems. Instead we evaluated each of our agents individually then calculated their overall mean to calculate the metrics for the entire agentic system.

The tables below give the values obtained as calculations of various metrics and also the rogue score (calculated specifically for the market analysis agent)

\begin{table}[hbt!]
\centering
\caption{Rouge-1 for Market Trend Analysis}
\begin{tabular}{|c|c|c|c|}
\hline
\textbf{Models} & \textbf{Precision} & \textbf{Recall} & \textbf{F-Score} \\
\hline
llama3-8b-7192  & 0.4204  & 0.1316 & 0.2266 \\ 
\hline
llama3-70b-8192 & 0.4368  & 0.2698 & 0.3084 \\
\hline
gemma-7b-it     & 0.4526  & 0.2784 & 0.3184 \\
\hline
gemma2-9b-it    & 0.4850  & 0.5250 & 0.5042 \\
\hline
\end{tabular}
\end{table}

\begin{table}[hbt!]
\centering
\caption{Rouge-2 for Market Trend Analysis}
\begin{tabular}{|c|c|c|c|}
\hline
\textbf{Models} & \textbf{Precision} & \textbf{Recall} & \textbf{F-score} \\
\hline
llama3-8b-7192  & 0.0936 & 0.0472 & 0.0588 \\ 
\hline
llama3-70b-8192 & 0.1316 & 0.1086 & 0.1093 \\
\hline
gemma-7b-it     & 0.1074 & 0.0581 & 0.0692 \\
\hline
gemma2-9b-it    & 0.098  & 0.1067 & 0.1025 \\
\hline
\end{tabular}
\end{table}

\begin{table}[hbt!]
\centering
\caption{Rouge-L for Market Trend Analysis}
\begin{tabular}{|c|c|c|c|}
\hline
\textbf{Models} & \textbf{Precision} & \textbf{Recall} & \textbf{F-score} \\
\hline
llama3-8b-7192  & 0.1978  & 0.1209 & 0.1385\\
\hline
llama3-70b-8192 & 0.2266  & 0.1919 & 0.1928 \\
\hline
gemma-7b-it     & 0.2082  & 0.1228 & 0.1428 \\
\hline
gemma2-9b-it    & 0.1731   & 0.1873 & 0.1799 \\
\hline
\end{tabular}
\end{table}

We used ROUGE score for evaluating or market analysis agent because the market analysis agent analyses a large pool of resources available on the internet, gets the information and generates a concise summary of the information obtained. Based on our own dataset of prompts and responses, we tried to evaluate how much the two summaries matched, hence we used the ROUGE scores.

Table 3, Table 4, and Table 5 give the ROUGE scores of the models that we tested.

We also tested Mixtral-8x7B-32768 and found that the model was not able to make a web search using the web search tool to get the required content. Gemma2-9b-it had longer response times when using it for market analysis and information retrieval, at times it ran out of iteration limit

\section{\uppercase{Conclusions}}
\label{sec:conclusion}

In this work, we propose the idea of using multi-modal AI agents for product recommendations. Unlike other recommendation systems which use mainly text data and user history to give product recommendations. Our method takes into account multi-modal data like images. to give recommendations aligned to user preferences. Moreover, we developed a web interface to demonstrate the capabilities of our multi agent system which helps the researchers visualize how the agents work collaboratively to solve a complex problem and also helps the user in better user interaction.

\bibliographystyle{apalike}
{\small
\bibliography{example}}

\section*{\uppercase{Acknowledgements}}

Authors acknowledge the support of Veermata Jijabai Technological Institute, Mumbai, India for the successful completion of this research

\end{document}